\documentclass{article}
\usepackage{spconf,amsmath,graphicx}
\usepackage{multirow}
\usepackage{amsfonts}
\usepackage{amsmath}
\usepackage{graphicx}
\usepackage{subcaption}
\usepackage{array}
\usepackage{hyperref}
\usepackage[normalem]{ulem}

\title{Noisy-ArcMix: Additive Noisy Angular Margin Loss \\combined with Mixup for Anomalous Sound Detection}
\name{Soonhyeon Choi, Jung-Woo Choi \thanks{This study was supported by LIG Nex1 (No. G01220632), and BK21 FOUR program through the National Research Foundation (NRF) funded by the Ministry of Education of Korea.}}
\address{School of Electrical Engineering, KAIST, Daejeon, South Korea}
\begin{document}
\newcommand{\meq}[2]{\vspace{0.5\baselineskip}\begin{equation}\begin{alignedat}{3}#1\end{alignedat}\label{#2}\end{equation}} 

\newcommand{\aln}[1]{\begin{alignat}{3}#1\end{alignat}}
\newcommand{\alnstar}[1]{\begin{alignat*}{3}#1\end{alignat*}}

\newcommand{\lrp}[1]{\left(#1\right)}
\newcommand{\lrb}[1]{\left\{#1\right\}}
\newcommand{\lrsb}[1]{\left[#1\right]}
\newcommand{\abs}[1]{\left|#1\right|}
\newcommand{\norm}[1]{\left\lVert#1\right\rVert}
\newcommand{\lrvert}[1]{\left|#1\right|}
\newcommand{\lrVert}[1]{\left\lVert#1\right\rVert}
\newcommand{\real}[1]{\mathcal{R}e\left\{#1\right\}}
\newcommand{\imag}[1]{\mathcal{I}m\left\{#1\right\}}
\newcommand{\expt}[1]{E\left\{#1\right\}}

\def\where{\text{ where }}
\def\ba{{\mathbf{a}}}  \def\bb{{\mathbf{b}}}  \def\bc{{\mathbf{c}}} \def\bd{{\mathbf{d}}}
\def\bee{{\mathbf{e}}} \def\bff{{\mathbf{f}}} \def\bg{{\mathbf{g}}} \def\bh{{\mathbf{h}}}
\def\bi{{\mathbf{i}}}  \def\bj{{\mathbf{j}}}  \def\bk{{\mathbf{k}}} \def\bl{{\mathbf{l}}}
\def\bmm{{\mathbf{m}}}  \def\bn{{\mathbf{n}}}  \def\bo{{\mathbf{o}}} \def\bp{{\mathbf{p}}}
\def\bq{{\mathbf{q}}}  \def\br{{\mathbf{r}}}  \def\bs{{\mathbf{s}}} \def\bt{{\mathbf{t}}}
\def\bu{{\mathbf{u}}}  \def\bv{{\mathbf{v}}}  \def\bw{{\mathbf{w}}} \def\bx{{\mathbf{x}}}
\def\by{{\mathbf{y}}}  \def\bz{{\mathbf{z}}}
\def\bA{{\mathbf{A}}} \def\bB{{\mathbf{B}}} \def\bC{{\mathbf{C}}} \def\bD{{\mathbf{D}}}
\def\bE{{\mathbf{E}}} \def\bF{{\mathbf{F}}} \def\bG{{\mathbf{G}}} \def\bH{{\mathbf{H}}}
\def\bI{{\mathbf{I}}} \def\bJ{{\mathbf{J}}} \def\bK{{\mathbf{K}}} \def\bL{{\mathbf{L}}}
\def\bM{{\mathbf{M}}} \def\bN{{\mathbf{N}}} \def\bO{{\mathbf{O}}} \def\bP{{\mathbf{P}}}
\def\bQ{{\mathbf{Q}}} \def\bR{{\mathbf{R}}} \def\bS{{\mathbf{S}}} \def\bT{{\mathbf{T}}}
\def\bU{{\mathbf{U}}} \def\bV{{\mathbf{V}}} \def\bW{{\mathbf{W}}} \def\bX{{\mathbf{X}}}
\def\bY{{\mathbf{Y}}} \def\bZ{{\mathbf{Z}}}
\def\bone{{\mathbf{1}}} \def\bzero{{\mathbf{0}}}
\def\btheta{{\boldsymbol{\theta}}} \def\bTheta{{\mathbf{\Theta}}}
\def\bsigma{{\boldsymbol{\sigma}}} \def\bSigma{{\mathbf{\Sigma}}}
\def\blambda{{\boldsymbol{\lambda}}} \def\bLambda{{\mathbf{\Lambda}}}
\def\bgamma{{\boldsymbol{\gamma}}} \def\bGamma{{\mathbf{\Gamma}}}
\def\bphi{{\boldsymbol{\phi}}} \def\bPhi{{\mathbf{\Phi}}}
\def\bpsi{{\boldsymbol{\psi}}} \def\bPsi{{\mathbf{\Psi}}}
\def\bpi{{\boldsymbol{\pi}}} \def\bPi{{\mathbf{\Pi}}}
\def\beps{{\boldsymbol{\epsilon}}}



\def\H{^\mathsf{H}}
\def\mH{^{-\mathsf{H}}}
\def\T{^\mathsf{T}}
\def\th{^{\mathrm{th}}}



\def\inv{{^{-1}}}

\def\vC{{\mathbb{C}}} \def\vR{{\mathbb{R}}}
\newcommand{\inR}[1]{\!\in\!\mathbb{R}^{#1}}
\newcommand{\inC}[1]{\!\in\!\mathbb{C}^{#1}}

\def\cA{{\mathcal{A}}} \def\cB{{\mathcal{B}}} \def\cC{{\mathcal{C}}} \def\cD{{\mathcal{D}}}
\def\cE{{\mathcal{E}}} \def\cF{{\mathcal{F}}} \def\cG{{\mathcal{G}}} \def\cH{{\mathcal{H}}}
\def\cI{{\mathcal{I}}} \def\cJ{{\mathcal{J}}} \def\cK{{\mathcal{K}}} \def\cL{{\mathcal{L}}}
\def\cM{{\mathcal{M}}} \def\cN{{\mathcal{N}}} \def\cO{{\mathcal{O}}} \def\cP{{\mathcal{P}}}
\def\cQ{{\mathcal{Q}}} \def\cR{{\mathcal{R}}} \def\cS{{\mathcal{S}}} \def\cT{{\mathcal{T}}}
\def\cU{{\mathcal{U}}} \def\cV{{\mathcal{V}}} \def\cW{{\mathcal{W}}} \def\cX{{\mathcal{X}}}
\def\cY{{\mathcal{Y}}} \def\cZ{{\mathcal{Z}}} \def\cz{{\mathcal{z}}}
\def\SFT{\mathrm{SFT}}

\def\meter{\mathrm{\,m}}

\newcommand{\dm}[1]{\color{red}[#1]\color{black}}
\newcommand{\dmr}[1]{\color{blue}[#1]\color{black}}
\newcommand{\rev}[1]{\color{magenta}#1\color{black}}

\newcommand*\xbar[1]{%
  \hbox{%
    \kern 0.1em
    \vbox{%
      \hrule height 0.5pt 
      \kern0.3ex
      \hbox{%
        \kern-0.0em
        \ensuremath{#1}%
        \kern-0.0em
      }%
    }%
    \kern 0.0em
  }
} 

%
\maketitle
\begin{abstract}
Unsupervised anomalous sound detection (ASD) aims to identify anomalous sounds by learning the features of normal operational sounds and sensing their deviations. Recent approaches have focused on the self-supervised task utilizing the classification of normal data, and advanced models have shown that securing representation space for anomalous data is important through representation learning yielding compact intra-class and well-separated intra-class distributions. However,
we show that conventional approaches often fail to ensure sufficient intra-class compactness and exhibit angular disparity between samples and their corresponding centers.
In this paper, we propose a training technique aimed at ensuring intra-class compactness and increasing the angle gap between normal and abnormal samples. Furthermore, we present an architecture that extracts features for important temporal regions, enabling the model to learn which time frames should be emphasized or suppressed. Experimental results demonstrate that the proposed method achieves the best performance giving $\mathbf{0.90}\%$, $\mathbf{0.83}\%$, and $\mathbf{2.16}\%$ improvement in terms of AUC, pAUC, and mAUC, respectively, compared to the state-of-the-art method on DCASE 2020 Challenge Task2 dataset. The source codes are available at \url{https://github.com/soonhyeon/Noisy-ArcMix}
\end{abstract}

\begin{figure*}[ht] 
\begin{subfigure}{0.24\textwidth}
\centering 
\includegraphics[width=\linewidth]{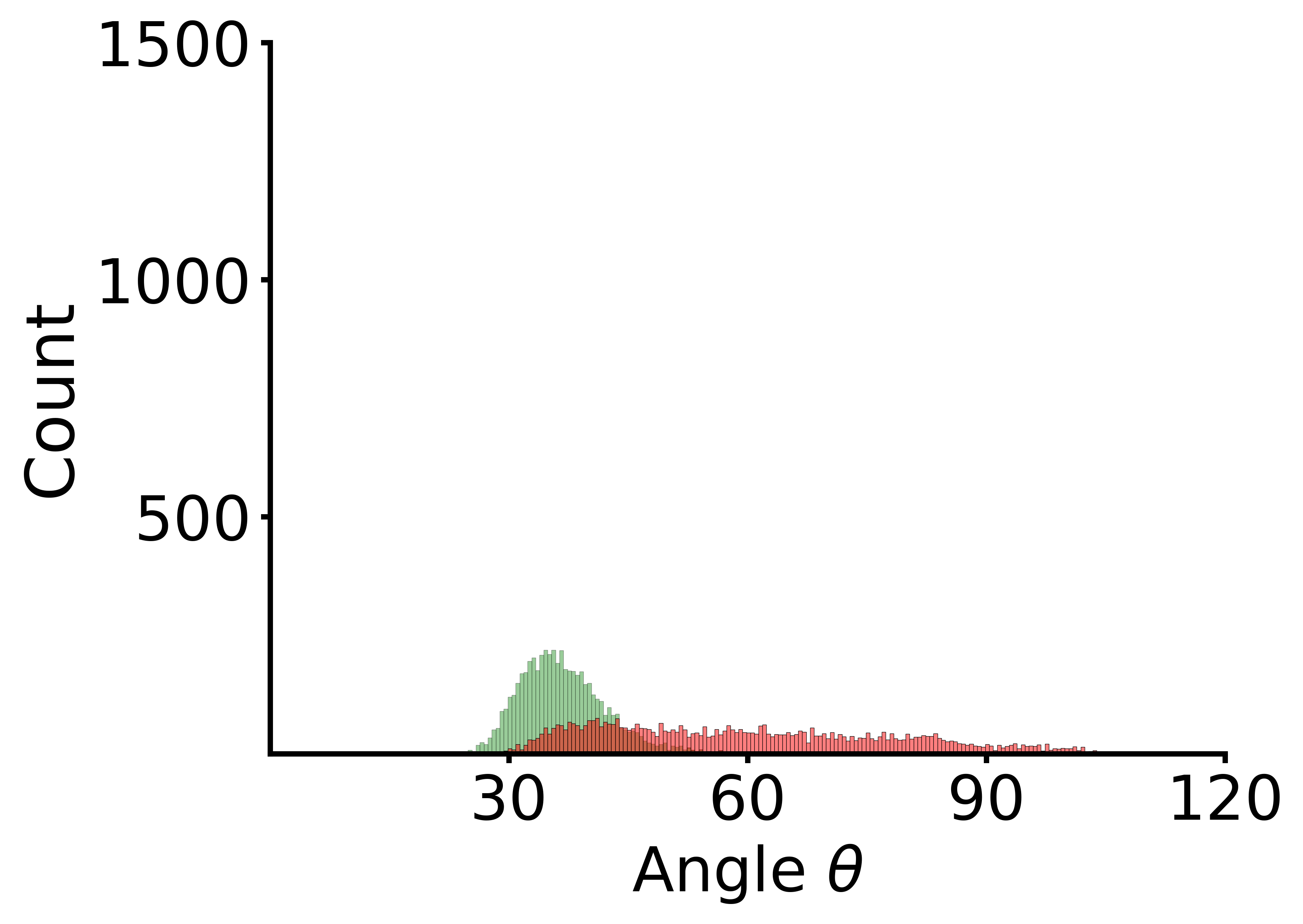}
\caption{Cross-Entropy}
\label{fig:cross_entropy}
\end{subfigure}
\begin{subfigure}{0.24\textwidth}
\centering 
\includegraphics[width=\linewidth]{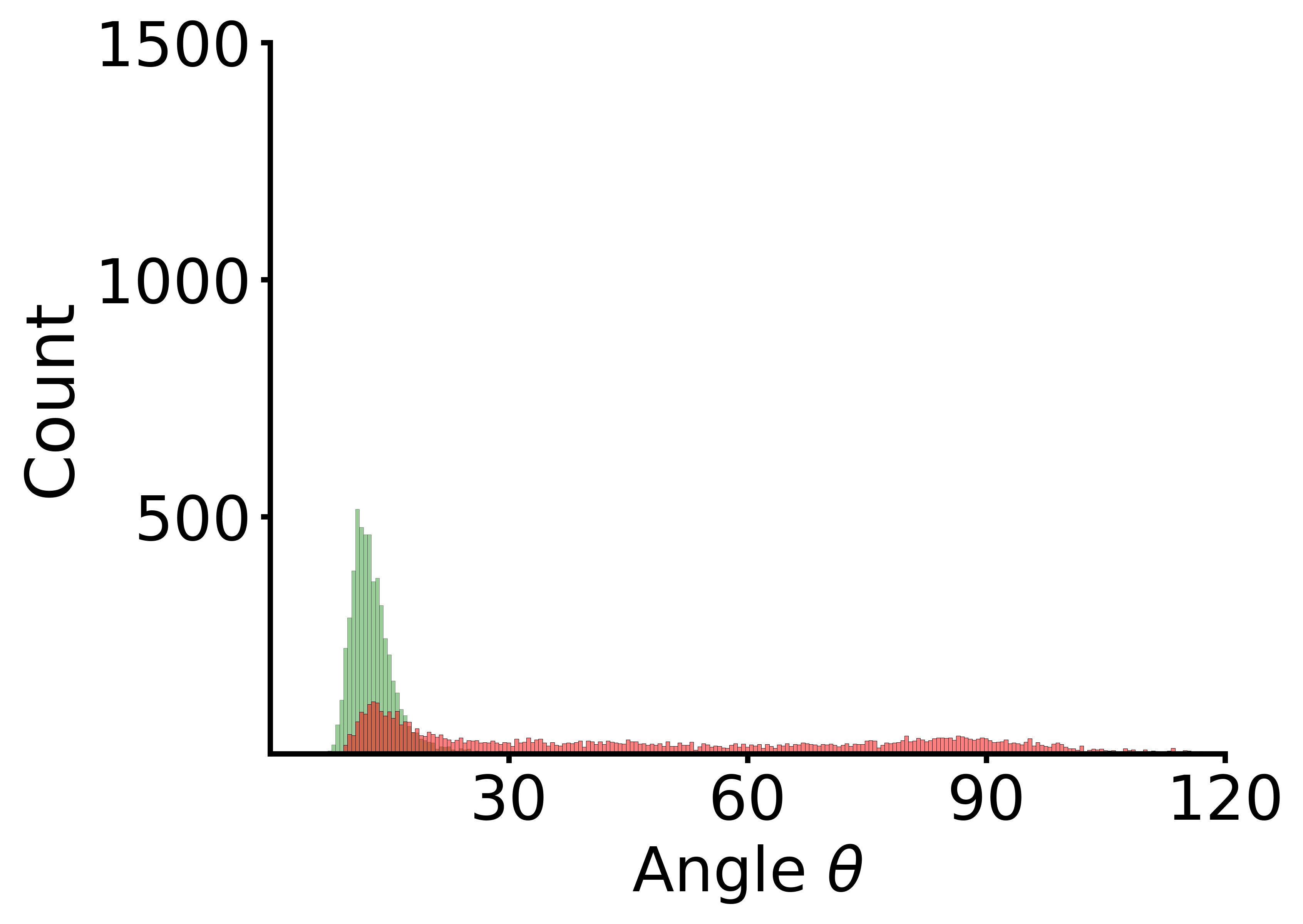}
\caption{ArcFace}
\label{fig:arcface}
\end{subfigure}
\begin{subfigure}{0.24\textwidth}
\centering 
\includegraphics[width=\linewidth]{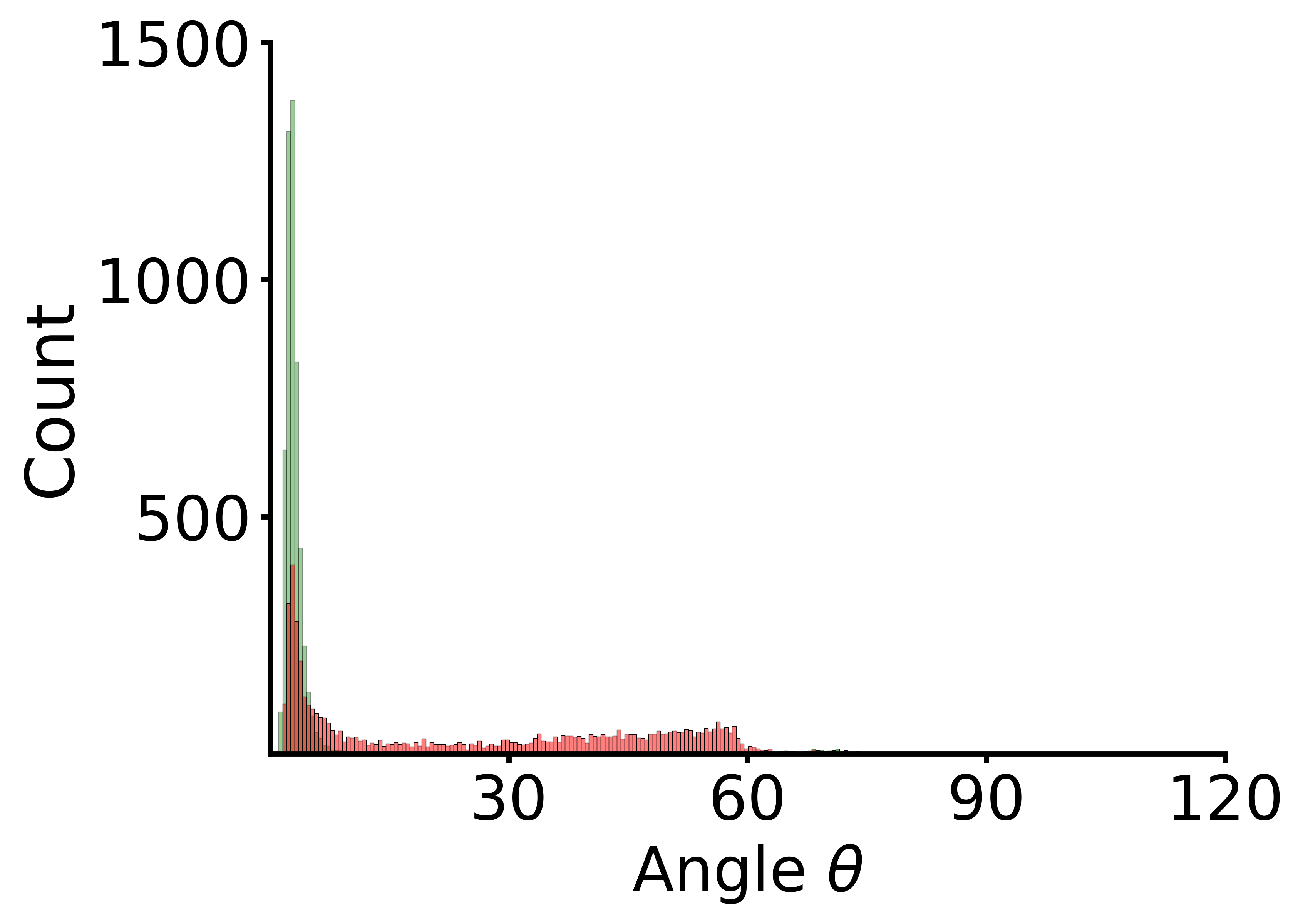}
\caption{ArcMix}
\label{fig:arcmix}
\end{subfigure}
\begin{subfigure}{0.24\textwidth}
\centering 
\includegraphics[width=\linewidth]{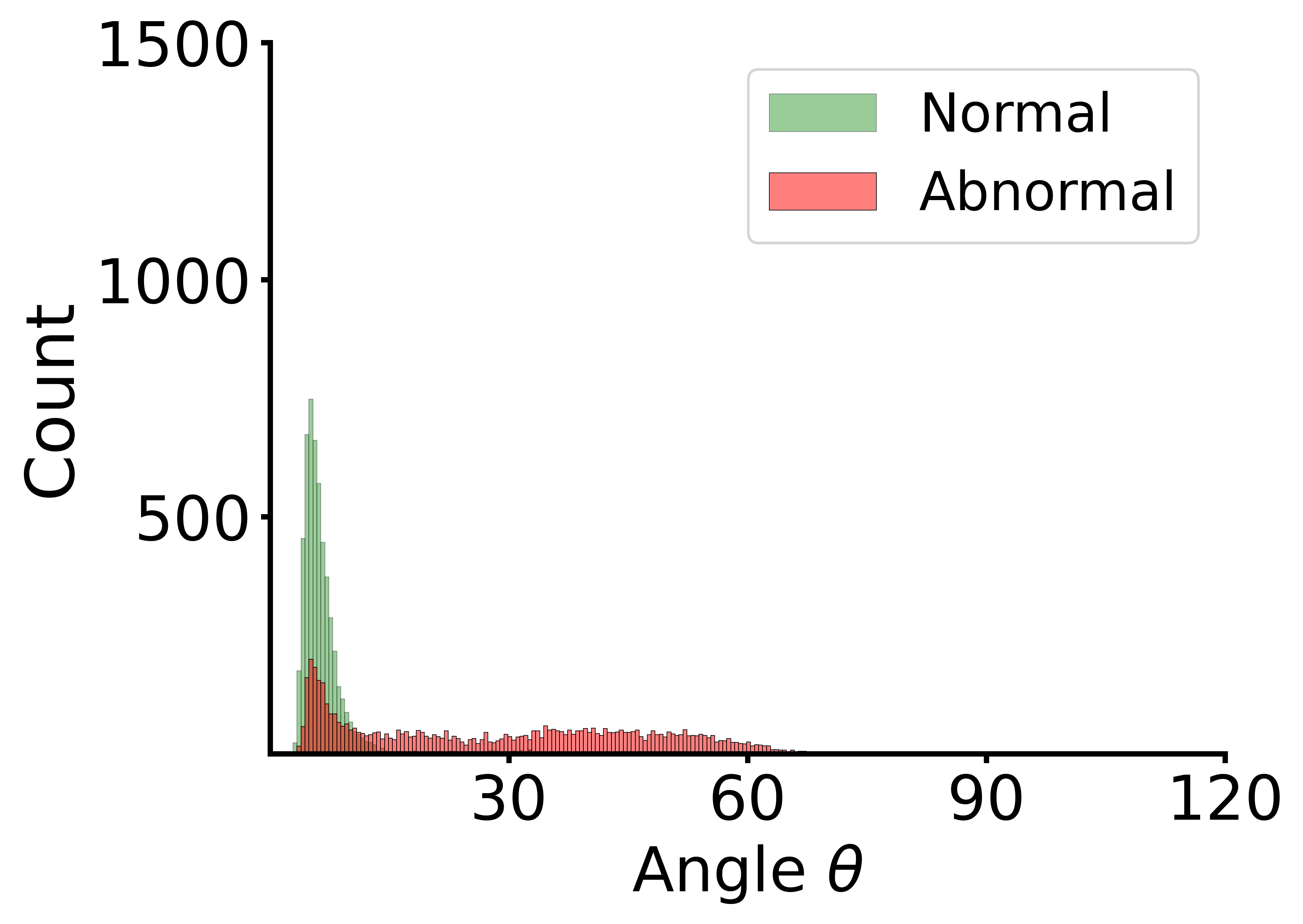}
\caption{Noisy-ArcMix}
\label{fig:noisy-arcmix}
\end{subfigure}
\caption{Distribution of angles between feature embeddings and corresponding learned class centers, for the models trained by (a) Cross-Entropy, (b) ArcFace, (c) ArcMix, and (d) Noisy-ArcMix. The results are derived from all machine types in the test data of DCASE 2020 Challenge Task 2 development dataset.}
\vspace{-0.5 cm}
\label{fig:fig2}
\end{figure*}

\section{Introduction}
\label{sec:intro}
In recent years, the increasing availability of machine sound data from various sources has spurred significant interest in ASD using machine learning techniques. Anomalous sounds refer to atypical sound patterns that deviate from the expected or normal acoustic behavior, which can arise from various factors such as equipment malfunctions, environmental disturbances, or unexpected events. 

Previous studies have focused on unsupervised learning of normal sounds without training on anomalies \cite{Koizumi_WASPAA2019_01, suefusa2020anomalous, giri2020self, dohi2021flow, Dohi2022-2}. In interpolation deep neural network (IDNN) \cite{suefusa2020anomalous}, a model is trained to minimize the reconstruction error, and the error is also used as an anomaly score. Since this approach is trained only on normal samples without anomalous data during training, the model can be easily overfitted to the training normal data. 
To mitigate the overfitting, Glow-Aff \cite{dohi2021flow} attempts to estimate the probability density of normal data using normalizing flow. In Glow-Aff, separate models are trained for individual machine types, such that each model can discriminate machine IDs of the data with the same machine type. However, the model trained from a small number of machine IDs cannot be discriminative enough and cannot be generalized well to unseen normal or abnormal data. 

In another promising approach, models designed to classify every machine type and machine ID have shown improved ASD performance \cite{giri2020self, dohi2021flow, liu2022anomalous, guan2023anomalous}. One such model, STgram-MFN \cite{liu2022anomalous}, extracts spectral and temporal features (i.e., Sgram and Tgram) and classifies the features through ArcFace loss \cite{deng2019arcface}. In particular, STgram-MFN facilitates the small intra-class distance and large inter-class distance through ArcFace. To consider global sample-to-class distance, ArcFace utilizes angular distance between the feature embedding of a data sample and the learned linear projection heads, i.e., learned class centers. ArcFace promotes intra-class compactness through the margin penalty directly added to the inter-class angles. 
On the other hand, CLP-SCF \cite{guan2023anomalous} utilizes contrastive learning in the pre-training stage, which reduces distances between pairs of feature embeddings from the same machine IDs and enlarges distances for pairs from different machine IDs. 
Although features learned by STgram-MFN and CLP-SCF can be more discriminative through the comparison to class centers or feature embeddings from other data samples, models are still trained using a limited number of machine IDs and normal training samples. Accordingly, the mingling of normal and anomalies unseen during the training can be induced at the test stage, which calls for the need to train with more versatile data.   




To overcome the limitation of ArcFace, we leverage mixup \cite{zhang2018mixup} that synthesizes samples near the normal data distribution, as well as corrupted labels, through the linear combination of two training samples and target labels, respectively. Mixup demonstrated that a more generalized model can be trained through the task of predicting corrupted labels from synthesized samples. The performance improvement over the ASD task was also reported \cite{wilkinghoff2021sub}. However, mixup and ArcFace cannot be simply combined since the way of imposing angular margins for the corrupted label prediction task is not straightforward. In this paper, we propose two approaches, ArcMix and Noisy-ArcMix, for utilizing angular margins on the mixup task.
We show that both ArcMix and Noisy-ArcMix significantly improve the compactness of intra-class distribution through the training with virtually synthesized samples near the normal data distribution. More importantly, we observed that the mingling effect between normal and anomalous samples can be reduced further by Noisy-ArcMix, which gains generalization ability through the use of inconsistent angular margins for the corrupted label prediction. 
In addition to Noisy-ArcMix, we introduce a new input feature, temporally attended log-Mel spectrogram (TAgram), derived from a temporal attention block in a similar fashion to \cite{woo2018cbam}. TAgram includes the temporal attention weights broadcasted to spectrogram features, which helps a model to focus on the important temporal regions for capturing crucial features. The performance of the proposed techniques is evaluated on the DCASE 2020 Challenge Task 2 dataset \cite{Koizumi_DCASE2020_01}, which demonstrates the effectiveness of our approach compared to the state-of-the-art methods across various machine types. 




\label{sec:format}

\useunder{\uline}{\ul}{}
\begin{table*}[ht]
\caption{Performance comparison in terms of AUC (\%) and pAUC (\%) on the test data of the development dataset. Boldface and underlined numbers denote the best and the top-3 results, respectively.} 
\scalebox{0.805}{
\begin{tabular}{cl|cccccccccccc|cc}
\hline
\multicolumn{2}{c|}{\multirow{2}{*}{Methods}} & \multicolumn{2}{c}{Fan}        & \multicolumn{2}{c}{Pump}        & \multicolumn{2}{c}{Slider}      & \multicolumn{2}{c}{Valve}       & \multicolumn{2}{c}{ToyCar}      & \multicolumn{2}{c|}{ToyConveyor} & \multicolumn{2}{c}{Average}  \\ \cline{3-16} 
\multicolumn{2}{c|}{}                         & AUC           & pAUC           & AUC            & pAUC           & AUC            & pAUC           & AUC            & pAUC           & AUC            & pAUC           & AUC            & pAUC           & AUC            & pAUC        \\ \hline
\multicolumn{2}{c|}{IDNN \cite{suefusa2020anomalous}}                     & 67.71         & 52.90           & 73.76          & 61.07          & 86.45          & 67.58          & 84.09          & 64.94          & 78.69          & 69.22          & 71.07          & 59.7           & 76.96          & 62.57       \\
\multicolumn{2}{c|}{MobileNetV2 \cite{giri2020self}}              & 80.19         & 74.40           & 82.53          & 76.50           & 95.27          & 85.22          & 88.65          & 87.98          & 87.66          & 85.92          & 69.71          & 56.43          & 84.34          & 77.74       \\
\multicolumn{2}{c|}{Glow-Aff \cite{dohi2021flow}}                  & 74.90          & 65.30           & 83.40           & 73.80           & 94.60           & 82.80           & 91.40           & 75.00             & 92.2           & 84.10           & 71.50           & 59.00             & 85.20           & 73.90        \\
\multicolumn{2}{c|}{STgram-MFN(ArcFace) \cite{liu2022anomalous}}      & 94.04   & 88.97    & 91.94    & 81.75    & {\ul 99.55}    & {\ul 97.61}    & {\ul 99.64}    & 98.44    & 94.44    & {\ul 87.68}    & {\ul 74.57}    & {\ul 63.6}     & 92.36    & 86.34 \\
\multicolumn{2}{c|}{CLP-SCF \cite{guan2023anomalous}}                  & {\ul 96.98}   & {\ul 93.23}    & {\ul 94.97} & \textbf{87.39} & \textbf{99.57} & \textbf{97.73} & {\ul 99.89}    & {\ul 99.51}    & {\ul 95.85}    & \textbf{90.19} & {\ul 75.21}    & {\ul 62.79}    & {\ul 93.75}    & {\ul 88.48} \\ \hline

\multicolumn{2}{c|}{TASTgram-MFN(AMix)}  & {\ul 96.43} & {\ul 94.52} & {\ul 93.12}    & {\ul 85.19}    & 99.08    & 95.23    & 99.22 & {\ul 98.83} & {\ul 95.54} & 87.55    & 70.89 & 61.30 & {\ul 92.38} & {\ul 87.10} \\

\multicolumn{2}{c|}{\textbf{TASTgram-MFN(NAMix)}}  & \textbf{98.32} & \textbf{95.34} & \textbf{95.44}    & {\ul 85.99}    & {\ul 99.53}    & {\ul 97.50}    & \textbf{99.95} & \textbf{99.74} & \textbf{96.76} & {\ul 90.11}    & \textbf{77.90} & \textbf{67.15} & \textbf{94.65} & \textbf{89.31} \\ \hline
\end{tabular}
}
\label{tab:tab1}
\end{table*}

\section{The Proposed Method}

\begin{table*}[t]
\caption{
Performance comparison in terms of AUC (\%) and pAUC (\%) for STgram and TASTgram architectures trained by ArcFace and Noisy-ArcMix.
}
\scalebox{0.805}{
\begin{tabular}{cl|cccccccccccc|cc}
\hline
\multicolumn{2}{c|}{\multirow{2}{*}{Methods}} & \multicolumn{2}{c}{Fan}        & \multicolumn{2}{c}{Pump}        & \multicolumn{2}{c}{Slider}   & \multicolumn{2}{c}{Valve}       & \multicolumn{2}{c}{ToyCar}      & \multicolumn{2}{c|}{ToyConveyor} & \multicolumn{2}{c}{Average}  \\ \cline{3-16} 
\multicolumn{2}{c|}{}                         & AUC           & pAUC           & AUC            & pAUC           & AUC            & pAUC        & AUC            & pAUC           & AUC            & pAUC           & AUC             & pAUC           & AUC            & pAUC        \\ \hline
\multicolumn{2}{c|}{STgram-MFN(ArcFace)}          & 94.04         & 88.97          & 91.94          & 81.75          & {\ul 99.55}          & {\ul 97.61} & 99.64          & 98.44          & 94.44          & 87.68          & {\ul 74.57}           & {\ul 63.60}           & 92.36          & 86.34       \\
\multicolumn{2}{c|}{TASTgram-MFN(ArcFace)}       & {\ul 95.46}         & {\ul 89.82}          & {\ul 93.41}          & {\ul 85.24}          & \textbf{99.78} & \textbf{98.81} & {\ul 99.71}    & {\ul 98.48}    & {\ul 95.78}    & {\ul 90.19} & 73.16     & 63.17    & {\ul 92.88}          & {\ul 87.62} \\
\multicolumn{2}{c|}{STgram-MFN(NAMix)}     & {\ul 97.2}          & {\ul 93.56}          & {\ul 94.89} & \textbf{87.07} & 99.51          & 97.44       & {\ul 99.88}          & {\ul 99.38}          & {\ul 96.55} & \textbf{90.41}    & {\ul 77.57}     & {\ul 64.54}    & {\ul 94.27}    & {\ul 88.73} \\
\multicolumn{2}{c|}{TASTgram-MFN(NAMix)}  & \textbf{98.32} & \textbf{95.34}    & \textbf{95.44}    & {\ul 85.99}    & {\ul 99.53}          & {\ul 97.50}       & \textbf{99.95} & \textbf{99.74} & \textbf{96.76}    & {\ul 90.11}    & \textbf{77.90}  & \textbf{67.15} & \textbf{94.65} & \textbf{89.31} \\ \hline
\end{tabular}
}
\vspace{-0.3cm}
\label{tab:tab2}
\end{table*}


\subsection{Noisy-ArcMix}
ArcFace \cite{deng2019arcface} increases intra-class compactness and inter-class discrepancy by incorporating an additive angular margin. The ArcFace loss for an input vector $\bx$ can be expressed in a vector form as 
\begin{align} \label{eq:arcface}
\begin{aligned}
\mathcal{L}_{\text{AF}}(\bx,\by) = - \by\T \log \frac{e^{s \cos{(\btheta + m \by)}}}{\sum^K_{k=1} e^{s \cos{(\theta_k + m y_k)}}}, 
\end{aligned}
\end{align}
where $\by=[y_1,~\cdots,~ y_K]\T$ is a one-hot vector for indicating the true label of $\bx$.    
The angular vector denoted by $\boldsymbol{\theta} = [\theta_1,~ \cdots,~ \theta_K]\T$ consists of the angles $\theta_k = \arccos (\bw_k\T \bh)$ between the normalized feature embedding $\bh \inR{d \times 1}$ from the feature extractor and the normalized linear projection head $\bw_k \inR{d \times 1}$ for class $k$. ArcFace introduces an additive angular margin penalty $m$ between $\bh$ and $\bw_k$ only for the true label $y_k = 1$ to promote compact intra-class distribution. The hyperparameter $s$ is a scaling parameter for the modified logit $\cos(\btheta + m\by)$. 
The margin is desirable to realize compact intra-class data distribution and to secure more space for anomalies unseen during training. 
As an example, Fig.\:\ref{fig:cross_entropy} and \ref{fig:arcface} present the distribution of angles $\theta$ to the ground-truth class center obtained from models trained without margin (cross-entropy) and with angular margins (ArcFace), respectively (details of the dataset are described in Sec. 3.1). The support of normal data distribution is more compact for ArcFace compared to cross-entropy. However, ArcFace without data augmentation still exhibits angular bias for both normal and abnormal samples, indicating insufficient intra-class sample compactness.


To address this issue, we attempt to combine ArcFace with the mixup training \cite{zhang2018mixup} that synthesizes new samples near the normal data distribution through the asymmetric mixing of normal samples. 
\begin{equation} \label{eq:mixup}
\begin{split}
{\mathbf{x}}^{ij} &= \lambda \mathbf{x}^i + (1-\lambda) \mathbf{x}^j,\\
\end{split}
\end{equation}
Here, $i \in \{1,~\cdots,~B\}$ and $j \in \text{Shuffle}\{1,~\cdots,~B\}$ denote the indices of data $\bx$ in the same minibatch of size $B$. The mixup weight $\lambda \sim Beta(\alpha, \alpha)$ is sampled from the Beta distribution with $\alpha=0.5$ to ensure that the input is close to either 0 or 1. Accordingly, the class label corresponding to the data with a higher mixup weight acts as the target label, and the other with a smaller weight becomes the noise label. To combine the mixup training with ArcFace, we first define the ArcMix loss as the weighted sum of two ArcFace losses for $\by^i$ and $\by^j$. That is,
\begin{equation} \label{eq:idgen}
\begin{split}
\mathcal{L}_{\text{AMix}}(\bx^{ij}\!,\!\by^i\!,\!\by^j)\!=\!\lambda \mathcal{L}_{\text{AF}}(\bx^{ij}\!,\!\by^i)\!+\!(1\!-\!\lambda) \mathcal{L}_{\text{AF}}(\bx^{ij}\!,\!\by^j),
\end{split}
\end{equation}
where $\bx^{ij}$ denotes the input constructed from mixup data \eqref{eq:mixup}.
Since the weight $\lambda$ is close to either 0 or 1, the ArcMix loss is dominated by the ArcFace loss of $i$ or $j$. Accordingly, the feature embedding of the generated sample $\bx^{ij}$ is pulled towards the center of the intra-class distribution, yielding some robustness against unseen normal samples near the normal data distribution. However, due to the increased intra-class compactness, abnormal samples can also be mingled with the intra-class distribution. To mitigate this problem, we propose the Noisy-ArcMix loss defined as
\begin{equation} \label{eq:noArcMix}
\begin{split}
\mathcal{L}_{\text{NAMix}}(\bx^{ij}\!,\by^{ij}) = -{\mathbf{y}^{ij}}\T \log \frac{e^{s \cos{(\boldsymbol{\theta} + m \mathbf{y}^i)}}}{\sum^K_{k=1} e^{s \cos{({\theta}_k + m y^i_k)}}}.
\end{split}
\end{equation}
Here, $\mathbf{y}^{ij} = \lambda \mathbf{y}^i + (1-\lambda) \mathbf{y}^j$ denotes the corrupted label. In this way, we aim to predict the corrupted label from the softmax probability of ArcFace by posing the margin penalty only to the data $\bx^i$. When $\lambda$ is large, the margin penalty promotes the compact intra-class distribution as in ArcFace. However, when $\lambda$ is small, the margin applied to the data $\bx^i$ with a smaller mixup weight emphasizes its angular distance to the center of the noise class. Accordingly, a model trained to minimize \eqref{eq:noArcMix} promotes the prediction of smaller probability $\lambda\by^i$ among $\by^{ij}$. This can enhance the model's sensitivity to discriminate the small perturbation of normal data, which is in line with the vicinal risk minimization principle \cite{chapelle2000vicinal}.




\begin{figure}[ht] 
\begin{center}
\includegraphics[width=\linewidth]{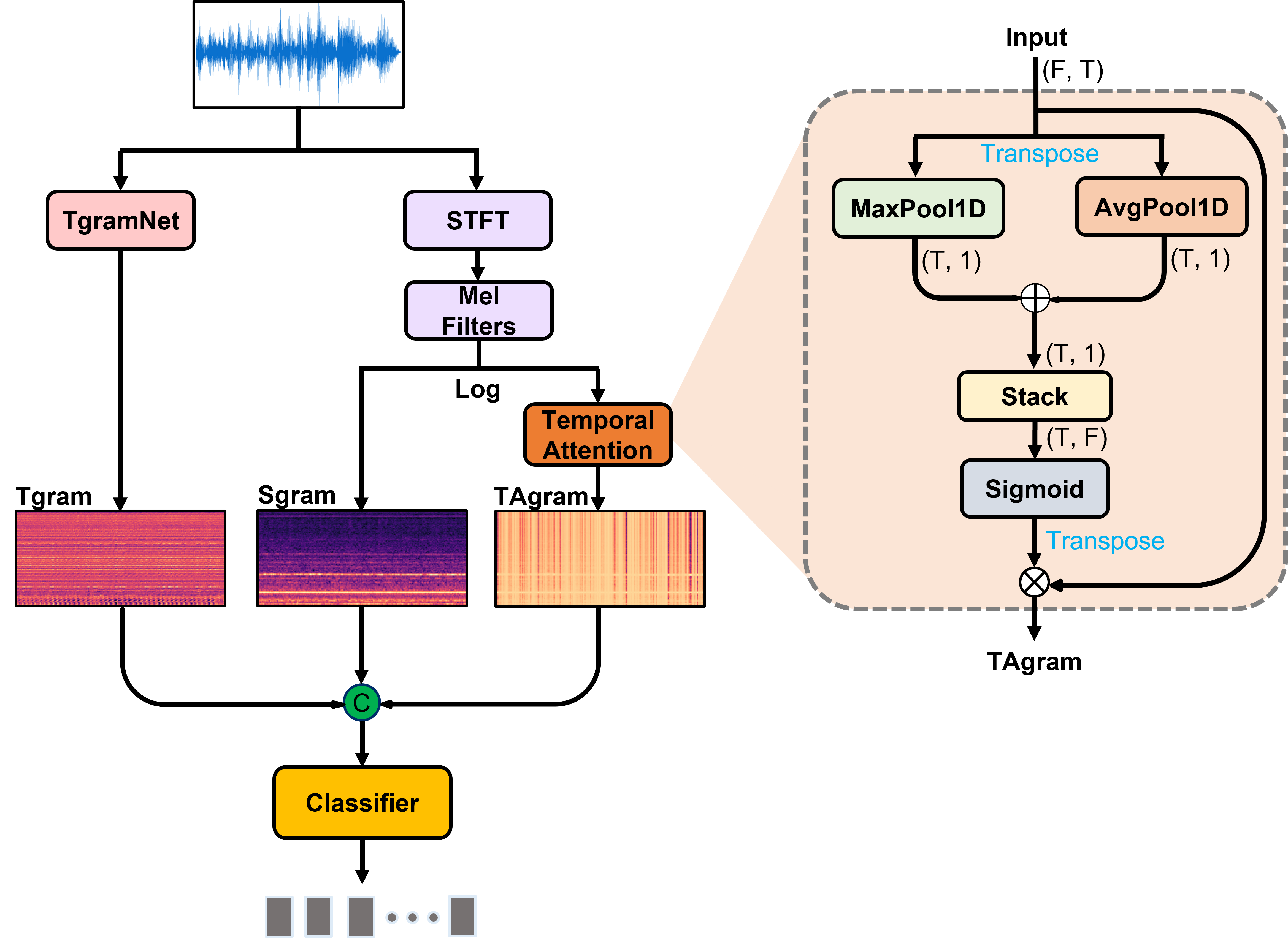}
\end{center}
\caption{Input architecture of TASTgramNet. The temporal feature (Tgram) is concatenated with the log-Mel spectrogram (Sgram) and the temporally attended feature (TAgram) from the temporal attention block.}
\label{fig:TASTnet}
\end{figure}

\subsection{TASTgramNet Architecture}
The STgram architecture \cite{liu2022anomalous} converts a single-channel input audio signal $\bx_{wav} \inR{1 \times D}$ of length $D$ to a log-Mel spectrogram $\bx_{mel} \inR{F \times T}$ with the number of mel bins $F$ and the number of time frames $T$. STgram also incorporates Tgram to extract temporal features from the raw wave through CNNs ($\bx_T = \textit{TgramNet}(\bx_{wav})$) \cite{liu2022anomalous}. In addition to these features, we propose to use features processed by temporal attention (TAgram) to emphasize important temporal regions. As illustrated in Fig.\:\ref{fig:TASTnet}, we generate a temporal attention map from the max-pooling and average-pooling operations along the spectral axis of the log-Mel spectrogram.  
\begin{equation} \label{eq:pool}
\begin{split}
Pool(\bx_{mel}) = AvgPool(\bx\T_{mel}) \oplus MaxPool(\bx\T_{mel}),
\end{split}
\end{equation}
where $\oplus$ denotes element-wise addition. After computing each pooling operation, a sigmoid function is applied to determine which parts should be emphasized or suppressed. The temporal attention map generated in this way is subsequently broadcasted along the temporal axis as
\begin{equation} \label{eq:tagram}
\begin{split}
\bx_{TA} = \sigma {(Pool(\bx\T_{mel}))} \otimes \bx_{mel},
\end{split}
\end{equation}
where $\otimes$ represents element-wise multiplication and $\sigma(\cdot)$ denotes the sigmoid function. 
Finally, $\bx_{TAST}$ is obtained by concatenating the log-Mel spectrogram $\bx_{mel}$, Tgram $\bx_{T}$ and TAgram $\bx_{TA}$ along the channel dimension.
\begin{equation} \label{eq:TASTgram}
\begin{split}
\bx_{TAST} = Concat(\bx_{TA}, \bx_{mel}, \bx_T)
\end{split}
\end{equation}
For ArcMix and Noisy-ArcMix, $\bx_{TAST}^{ij}$ is synthesized using \ref{eq:mixup} and then fed into the classifier (Mobile-FaceNet \cite{chen2018mobilefacenets}) as an input to learn representations.

\begin{table}[t]
\centering
\caption{
Performance comparison in terms of mAUC (\%) on the test data of the development dataset.
}
\scalebox{0.75}{
\begin{tabular}{c|ccc}
\hline
Method      & \begin{tabular}[c]{@{}c@{}}STgram-MFN\\ (ArcFace)\end{tabular} & CLP-SCF        & \textbf{\begin{tabular}[c]{@{}c@{}}TASTgram-MFN\\ (Noisy-ArcMix)\end{tabular}} \\ \hline
Fan         & 81.39                                                          & 88.27          & \textbf{92.67}                                                          \\
Pump        & 83.48                                                          & 87.27          & \textbf{91.17}                                                          \\
Slider      & 98.22                                                          & \textbf{98.28} & 97.96                                                                   \\
Valve       & 98.83                                                          & 99.58          & \textbf{99.89}                                                          \\
ToyCar      & 83.07                                                          & 86.87          & \textbf{88.81}                                                          \\
ToyConveyor & 64.16                                                          & 65.46          & \textbf{68.18}                                                          \\ \hline
Average     & 84.86                                                          & 87.62          & \textbf{89.78}                                                          \\ \hline
\end{tabular}
}
\vspace{-0.3 cm}
\label{tab:tab3}
\end{table}

\section{Experimental Results}
\label{sec:pagestyle}

\subsection{Experimental setup}

\textbf{\textit{Dataset}}
To evaluate our approach, we utilized the DCASE 2020 challenge Task 2 development dataset, which includes MIMII \cite{Purohit_DCASE2019_01} and ToyADMOS \cite{Koizumi_WASPAA2019_01} dataset, along with an additional dataset. The MIMII dataset comprises four machine types (Fan, Pump, Slider, and Valve) while the ToyADMOS dataset includes two machine types (ToyCar and ToyConveyor). The ToyCar dataset consists of six different machines and each of the other machine types has seven different machines. Therefore, a total of $41$ different machines are represented with approximately $10$ seconds of audio signals. The training data from the development dataset combined with the additional dataset were used for training. For evaluation, the test data from the development dataset was employed. 

\setlength{\parskip}{3pt}
\noindent\textbf{\textit{Implementation Details}} 
We trained the proposed TASTgram-MFN for the classification of $41$ labels derived from machine types and IDs \cite{liu2022anomalous, guan2023anomalous}. The process of generating Sgram $\mathbf{x}_{mel}$ and Tgram $\mathbf{x}_T$ is identical to that of \cite{liu2022anomalous}, and both have dimensions of $128 \times 313$. We employed ArcFace as the anomaly score in the same way as in \cite{liu2022anomalous}. For the Noisy-ArcMix, parameters were set as $\alpha=0.5$, $m=0.7$, and $s=30$. The network was trained using AdamW optimizer \cite{loshchilov2018decoupled} with a learning rate of $0.0001$, epochs of $300$, and a batch size of $64$. 

\noindent\textbf{\textit{Evaluation Metrics}}
For performance evaluation, we employed AUC, partial AUC (pAUC), and minimum AUC (mAUC). The pAUC is a metric that assesses the performance of the binary classification model within a specific range $[0, p]$ of false positive rates on the ROC curve. We set $p = 0.1$, as presented in \cite{liu2022anomalous, guan2023anomalous}. The mAUC represents the worst detection performance among the machines with different machine IDs, which was evaluated to examine detection stability \cite{dohi2021flow, liu2022anomalous, guan2023anomalous}.
\setlength{\parskip}{0pt}

\subsection{Performance comparison}
\label{sec:performance}
We compared ASD performance of the proposed approach with other methods in Table\:\ref{tab:tab1}. 
The results demonstrate that the proposed approach achieves the best performance in terms of average AUC and pAUC with improvements of $\mathbf{0.90}\%$ and $\mathbf{0.83}\%$ compared to the state-of-the-art, CLP-SCF.  The performance improvement is consistent across most machine types with only the exception for the slider data. Furthermore, the proposed approach also shows a significant improvement of $\mathbf{2.16}\%$ over CLP-SCF in mAUC (Table\:\ref{tab:tab3}), which also verifies its enhanced detection stability on various machine data.

In the distribution of angles to corresponding class centers (Fig.\:\ref{fig:fig2}), Cross-Entropy and ArcFace exhibit a substantial angular disparity between the samples and their respective centers. 
On the contrary, training with ArcMix (Fig.\:\ref{fig:arcmix}) shows enhanced inter-class compactness without bias, through generalization using synthesized training samples and margin penalty added to both classes $\by^i$ and $\by^j$. 
However, this also results in an undesirable intermingling effect between normal and abnormal samples. This artifact is reduced in Noisy-ArcMix (Fig.\:\ref{fig:noisy-arcmix}) which asymmetrically assigns the margin penalty to the angles corresponding to target or noise classes. This type of margin penalty finds a better compromise between inter-class compactness and discrimination of small perturbations from normal data distribution. As a result, the Noisy-ArcMix shows a significant improvement in ASD performance over ArcMix as presented in Table\:\ref{tab:tab1}.

Lastly, to demonstrate the efficacy of TASTgram and Noisy-ArcMix, we applied ArcFace and Noisy-ArcMix to STgram and TASTgram, respectively, and compared the performance in Table\:\ref{tab:tab2}. The comparison between STgram and TASTgram reveals that TASTgram outperforms STgram for both ArcFace and Noisy-ArcMix. Although most of the performance gain comes from the use of Noisy-ArcMix, the consistent performance improvement brought by TASTgram indicates that the temporal attention on the Sgram is beneficial for detecting anomalous regions in various data. 


\section{Conclusion}
\label{sec:typestyle}
To achieve compact inter-class distribution and improved discrimination for anomalous samples, we introduced Noisy-ArcMix, which combines the advantages of ArcFace and mixup. Noise-ArcMix was realized by applying an asymmetric angular margin penalty to the target and noise classes during the training with data synthesized by mixup. Furthermore, we utilized the temporally attended feature (TAgram) to make a model to focus on dominant temporal regions. We demonstrated that the proposed approach with Noise-ArcMix and TAgram can improve ASD performance and detection robustness across various machine sounds, accomplishing state-of-the-art performance on the DCASE 2020 Task 2 dataset. 


\label{sec:majhead}



\bibliographystyle{IEEEbib}
\bibliography{ref}

\begin{thebibliography}{10}

\bibitem{Koizumi_WASPAA2019_01}
Y.~Koizumi, S.~Saito, H.~Uematsu, N.~Harada, and K.~Imoto,
\newblock ``{ToyADMOS}: A dataset of miniature-machine operating sounds for
  anomalous sound detection,''
\newblock in {\em Proceedings of IEEE Workshop on Applications of Signal
  Processing to Audio and Acoustics ({WASPAA})}, November 2019, pp. 308--312.

\bibitem{suefusa2020anomalous}
K.~Suefusa, T.~Nishida, H.~Purohit, R.~Tanabe, T.~Endo, and Y.~Kawaguchi,
\newblock ``Anomalous sound detection based on interpolation deep neural
  network,''
\newblock in {\em Proceedings of IEEE International Conference on Acoustics,
  Speech and Signal Processing (ICASSP)}, Barcelona, Spain, 2020, IEEE, pp.
  271--275.

\bibitem{giri2020self}
R.~Giri, S.~Tenneti, F.~Cheng, K.~Helwani, U.~Isik, and A.~Krishnaswamy,
\newblock ``Self-supervised classification for detecting anomalous sounds,''
\newblock in {\em Proceedings of Detection and Classification of Acoustic
  Scenes and Events 2020 Workshop (DCASE)}, 2020, pp. 46--50.

\bibitem{dohi2021flow}
K.~Dohi, T.~Endo, H.~Purohit, R.~Tanabe, and Y.~Kawaguchi,
\newblock ``Flow-based self-supervised density estimation for anomalous sound
  detection,''
\newblock in {\em Proceedings of IEEE International Conference on Acoustics,
  Speech and Signal Processing (ICASSP)}, Toronto, ON, Canada, 2021, IEEE, pp.
  336--340.

\bibitem{Dohi2022-2}
K.~Dohi et~al.,
\newblock ``Description and discussion on {DCASE} 2022 challenge task 2:
  Unsupervised anomalous sound detection for machine condition monitoring
  applying domain generalization techniques,''
\newblock in {\em Proceedings of Detection and Classification of Acoustic
  Scenes and Events 2022 Workshop (DCASE2022)}, Nancy, France, 2022, pp. 1--5.

\bibitem{liu2022anomalous}
Y.~Liu, J.~Guan, Q.~Zhu, and W.~Wang,
\newblock ``Anomalous sound detection using spectral-temporal information
  fusion,''
\newblock in {\em Proceedings of IEEE International Conference on Acoustics,
  Speech and Signal Processing (ICASSP)}, Singapore, 2022, IEEE, pp. 816--820.

\bibitem{guan2023anomalous}
J.~Guan, F.~Xiao, Y.~Liu, Q.~Zhu, and W.~Wang,
\newblock ``Anomalous sound detection using audio representation with machine
  id based contrastive learning pretraining,''
\newblock in {\em Proceedings of IEEE International Conference on Acoustics,
  Speech and Signal Processing (ICASSP)}, Rhodes Island, Greece, 2023, IEEE,
  pp. 1--5.

\bibitem{deng2019arcface}
J.~Deng, J.~Guo, N.~Xue, and S.~Zafeiriou,
\newblock ``Arcface: Additive angular margin loss for deep face recognition,''
\newblock in {\em Proceedings of IEEE/CVF conference on computer vision and
  pattern recognition (CVPR)}, Long Beach, CA, USA, 2019, pp. 4690--4699.

\bibitem{zhang2018mixup}
H.~Zhang, M.~Cisse, Y.~Dauphin, and D.~Lopez-Paz,
\newblock ``mixup: Beyond empirical risk minimization,''
\newblock in {\em Proceedings of International Conference on Learning
  Representations (ICLR)}, Vancouver, BC, Canada, 2018.

\bibitem{wilkinghoff2021sub}
K.~Wilkinghoff,
\newblock ``Sub-cluster adacos: Learning representations for anomalous sound
  detection,''
\newblock in {\em Proceedings of International Joint Conference on Neural
  Networks (IJCNN)}, Shenzhen, China, 2021, IEEE, pp. 1--8.

\bibitem{woo2018cbam}
S.~Woo, J.~Park, J.~Lee, and I.~Kweon,
\newblock ``Cbam: Convolutional block attention module,''
\newblock in {\em Proceedings of European conference on computer vision
  (ECCV)}, Munich, Germany, 2018, pp. 3--19.

\bibitem{Koizumi_DCASE2020_01}
Y.~Koizumi et~al.,
\newblock ``Description and discussion on {DCASE}2020 challenge task2:
  Unsupervised anomalous sound detection for machine condition monitoring,''
\newblock in {\em Proceedings of Detection and Classification of Acoustic
  Scenes and Events 2020 Workshop (DCASE2020)}, November 2020, pp. 81--85.

\bibitem{chapelle2000vicinal}
O.~Chapelle, J.~Weston, L.~Bottou, and V.~Vapnik,
\newblock ``Vicinal risk minimization,''
\newblock in {\em Proceedings of the 13th International Conference on Neural
  Information Processing Systems (NIPS)}, Denver, CO, USA, 2000, pp. 395--401.

\bibitem{chen2018mobilefacenets}
S.~Chen, Y.~Liu, X.~Gao, and Z.~Han,
\newblock ``Mobilefacenets: Efficient cnns for accurate real-time face
  verification on mobile devices,''
\newblock in {\em Biometric Recognition: Chinese Conference, CCBR 2018, Urumqi,
  China}. Springer, 2018, pp. 428--438.

\bibitem{Purohit_DCASE2019_01}
H.~Purohit et~al.,
\newblock ``{MIMII Dataset}: Sound dataset for malfunctioning industrial
  machine investigation and inspection,''
\newblock in {\em Proceedings of Detection and Classification of Acoustic
  Scenes and Events 2019 Workshop ({DCASE2019})}, November 2019, pp. 209--213.

\bibitem{loshchilov2018decoupled}
I.~Loshchilov and F.~Hutter,
\newblock ``Decoupled weight decay regularization,''
\newblock in {\em Proceedings of International Conference on Learning
  Representations (ICLR)}, Vancouver, BC, Canada, 2018.

\end{thebibliography}

\end{document}